\documentclass[twocolumn,showpacs,preprintnumbers,amsmath,amssymb,superscriptaddress,footinbib]{revtex4}
\usepackage{graphicx}% Include figure files
\usepackage{dcolumn}% Align table columns on decimal point
\usepackage{bm}% bold math
\usepackage{graphicx}
\usepackage{xspace}
\usepackage{multirow}
\usepackage{natbib}
\usepackage{color}
\usepackage{epstopdf}

\newcommand{\bitwo}{Bi$_{2}$Sr$_2$CaCu$_2$O$_{8+x}$}
\newcommand{\bione}{Bi$_{2}$Sr$_2$CuO$_{6+x}$}
\newcommand{\ybco}{YBa$_2$Cu$_3$O$_{6+x}$}

\newcommand{\lsco}{La$_{2-x}$Sr$_x$CuO$_4$}

\begin{document}

\preprint{}

\title{Unusual photoemission resonances of oxygen-dopant induced states in \bitwo}

\author{P. Richard}\email{richarpi@bc.edu}
\affiliation{Department of Physics, Boston College, Chestnut Hill, Massachusetts 02467}
\author{Z.-H. Pan}
\affiliation{Department of Physics, Boston College, Chestnut Hill, Massachusetts 02467}
\author{M. Neupane}
\affiliation{Department of Physics, Boston College, Chestnut Hill, Massachusetts 02467}
\author{A. V. Fedorov}
\affiliation{Advanced Light Source, Lawrence Berkeley National Laboratory, Berkeley, California 94720}
\author{T. Valla}
\affiliation{Condensed Matter Physics and Materials Science Department, Brookhaven National Laboratory, Upton, New York 11973}
\author{P. D. Johnson}
\affiliation{Condensed Matter Physics and Materials Science Department, Brookhaven National Laboratory, Upton, New York 11973}
\author{G. D. Gu}
\affiliation{Condensed Matter Physics and Materials Science Department, Brookhaven National Laboratory, Upton, New York 11973}
\author{W. Ku}
\affiliation{Condensed Matter Physics and Materials Science Department, Brookhaven National Laboratory, Upton, New York 11973}
\author{Z. Wang}
\affiliation{Department of Physics, Boston College, Chestnut Hill, Massachusetts 02467}
\author{H. Ding}
\affiliation{Department of Physics, Boston College, Chestnut Hill, Massachusetts 02467}

\date{\today}% It is always \today, today,

\begin{abstract}

We have performed an angular-resolved photoemission study of underdoped, optimally doped and overdoped \bitwo\xspace samples using a wide photon energy range (15 - 100 eV). We report a small and broad non-dispersive A$_{1g}$ peak in the energy distribution curves whose intensity scales with doping. We attribute it to a local impurity state similar to the one observed recently by scanning tunneling spectroscopy and identified as the oxygen dopants. Detailed analysis of the resonance profile and comparison with the single-layered \bione\xspace suggest a mixing of this local state with Cu via the apical oxygens.
\end{abstract}

%\verb+\pacs{#1}+ command.

%Use showkeys class option if keyword
\maketitle

%\section{Introduction}
In contrast to pure systems, the carrier density of the CuO$_2$ planes in the cuprates can only be modified by creating nonstoichiometric defects in the parent compounds. For example, Bi$_2$Sr$_2$CaCu$_2$O$_{8+x}$ (Bi2212) is hole-doped mainly by the introduction of an extraneous amount of oxygens, which is estimated to be around x=0.16 per formula unit for the T$_c$ = 95 K optimally doped samples \cite{Presland}. Such dopants have long been suspected to induce local electronic disorder \cite{Pan1}. However, it is only recently that scanning tunneling spectroscopy (STS) measurements established their in-plane positions in the structure by correlating the STS spectra to the local electronic disorder as a function of doping \cite{Mcelroy}. In particular, the differential conductance at the oxygen dopant (O$_{\delta}$) site is characterized by a broad peak around -0.96 eV. This peak is observed in $\sim$ 8 \AA\xspace diameter regions and their number is found to be proportional to doping \cite{Mcelroy}. Surprisingly, larger superconducting gap magnitudes accompanied with the absence of coherent peaks are observed in these regions, suggesting that the O$_{\delta}$ are correlated to the CuO$_2$ planes electronic properties. Due to the importance of this issue, it is necessary to confirm the existence of that coupling and to investigate further the local impact of the O$_{\delta}$. Owing to its complementarity with STS, APRES is certainly an appropriate tool to investigate this issue. More importantly, the atomic selectivity of the resonant ARPES measurement provides unique and valuable insight concerning the wavefunction of the states of interest. When the incident photon energy is tunned to near an absorption edge of a specific element, a resonance (strong enhancement of photoelectron emission) usually occurs for certain valence spectral features that involve this element. This ``element-resolved" capability may shed light on the nature of wavefunctions at the O$_{\delta}$ site and its influence to local superconducting properties.

In this letter, we present an ARPES study of Bi2212 which reveals a small, broad and non-dispersing peak corresponding to a localized state analogous to the one reported recently by STS at -0.96 eV \cite{Mcelroy}. We show that the spectral weight of the corresponding ARPES peak increases with doping, thus reinforcing its relation to the presence of O$_{\delta}$. In order to characterize further this feature, we report measurements obtained for a wide photon energy range. The peak exhibits photon energy resonances around 50 eV and 75 eV. The latter, corresponding to the Cu3p$\rightarrow$3d edge, suggests a mixing of the localized state with Cu. Furthermore, the absence of resonance associated with the O2s$\rightarrow$2p transition implies that the state can only involve fully ionized oxygens. We show that the peak is likely to have A$_{1g}$ symmetry, which suggests a mixing with the apical oxygen O2p$_z$ and Cu3d$_{3z^2-r^2}$ orbitals. This interpretation is consistent with the lack of similar feature in Bi2201, in which the O2p$_z$-Cu3d$_{3z^2-r^2}$ is found $\sim$1 eV higher in binding energy according to our theoretical calculation.

%\section{Experiment}

High-quality single crystals of Bi2212 with various doping have been grown by traveling solvent floating zone method and subsequently annealed. For sake of clarity, we use in the text the notation XT to identify the doping level of the samples. In this shorten notation, X refers to underdoped (UD), optimally doped (OP) or overdoped (OD) samples, while T corresponds to the superconducting temperature. For example, OD71 means overdoped sample with T$_c$ = 71 K. ARPES spectra at 20 K were measured with photon energy from 28 eV to 100 eV using a Scienta SES-100 and the synchrotron beamline 12.001 of the Advanced Ligth Source, CA. 
For these measurements, the rotating capability of both the sample holder and the analyzer allowed us to record spectra with polarization parallel or perpendicular to $\Gamma$(0,0)-Y($\pi$,$\pi$) and M(0,$\pi$)-Y on the same samples. Measurements were also obtained at 10 K for the 15-23 eV photon energy range using a Scienta SES-2002 and the synchrotron beamline U13UB of the National Synchrotron Ligth Source, NY. The energy resolution is $\sim$ 10 - 40 meV for the photon energy range used in this study. Samples have been cleaved and measured \emph{in situ} in a vacuum better than $8\times10^{-11}$ $Torr$ on a flat (001) surface.

%\section{Results and discussion}

\begin{figure}[htbp]
\begin{center}
\includegraphics[width=8cm]{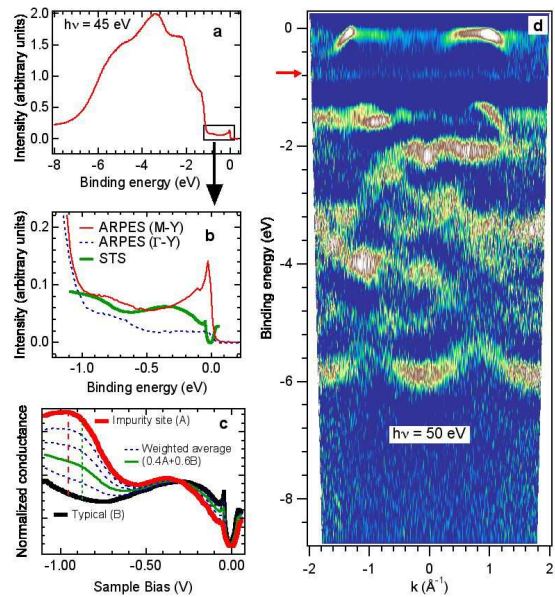}
\caption{\label{fig_pres}(Color online) a) k$_F$-centered EDC of an OD71 sample along M-Y. b) Zoom of the M-Y (thin plain line) and $\Gamma$-Y (dotted line) EDCs at k$_F$ compared to the weighted average of the STS typical (60 \%) and O$_{\delta}$ site (40 \%) spectra (thick line) from ref. \cite{Mcelroy}. c) Weighted averages (20\% steps) of the typical and O$_{\delta}$ site STS spectra from ref. \cite{Mcelroy}. The left vertical dashed line and right vertical dotted line correspond to the -0.96 eV STS peak and to the slope change from the weighted curve (third curve from the bottom), respectively. d) Second derivative intensity plot of an OP90 sample along $\Gamma$-M.}
\end{center}
\end{figure}

The complementarity of the ARPES and STS techniques suggests that ARPES spectra should exhibit fingerprints of the -0.96 eV STS peak associated with O$_{\delta}$. Besides normalization and matrix element effects, the space-integrated STS spectrum should correspond to the momentum-integrated ARPES one. As illustrated in Fig. \ref{fig_pres}a, a small and broad peak is revealed in the energy distribution curve (EDC) of an OD71 sample obtained at a photon energy of 45 eV within a 0.3 \AA$^{-1}$ momentum window centered at the k$_F$ point (we refer this as k$_F$-centered throughout the paper) of the anti-nodal direction (M-Y). A zoom of this EDC around -1.2 and 0.2 eV is given in Fig. \ref{fig_pres}b. In order to illustrate the effect of space averaging on the -0.96 eV STS peak and to reproduce the ARPES results, we show in Fig. \ref{fig_pres}c weighted averages (20\% steps) of the two STS curves digitized from Fig. 1(A) in ref. \cite{Mcelroy}. The third curve from the bottom (40\% typical spectrum+60\% impurity site spectrum \footnote{Assuming an amount of 0.11 O$_{\delta}$ per CuO$_2$ plane resulting from the doping of an overdoped sample (x = 0.22), this would correspond to impurity sites with size of $\sim$5 \AA\xspace radius.}) is reprodced in Fig. \ref{fig_pres}b. A good qualitative agreement is found between this curve and the OD71 k$_F$-centered EDCs measured along the nodal and anti-nodal directions. As the weight of the STS impurity state spectrum decreases and the spectrum evolves to the typical STS spectrum, the peak is smoothly suppressed and manifests its presence by a slope change rather than a well defined maximum. Even though the exact position of the slope change depends on the normalization procedure of the STS spectra, it is always observed at lower binding energies than the -0.96 eV peak. This is illustrated in Fig. \ref{fig_pres}c by the dotted line associated with the weighted STS curve slope change. Interestingly, the -0.8 eV ARPES peak does not show any dispersion, as illustrated by the second derivative intensity plot shown in Fig. \ref{fig_pres}d, in contrast to most of the other bands detected in the -7 to 0.3 eV energy range. This is consistent with the local impurity state origin of the feature.

%\newpage
 
\begin{figure}[htbp]
\begin{center}
\includegraphics[width=8cm]{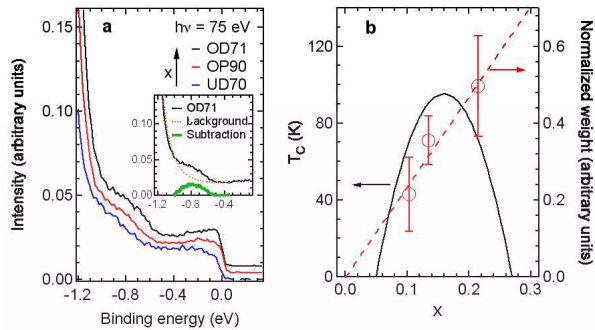}
\caption{\label{x}(Color online) a) k$_F$-centered EDCs of the UD70 (bottom), OP90 (middle) and OD71 (top) samples EDCs obtained along $\Gamma$-Y. Inset: background (dotted curve) associated with the OD71 EDC and the subtraction of the two curves (thick line). b) Doping dependence of the -0.8 eV peak spectral weight (circles), empirical relation (plain) between T$_c$ and the doping x given in ref. \cite{Presland} (T$_c^{\textrm{max}}$ = 95 K), and proportional relation between the spectral weight and x (dashed line).}
\end{center}
\end{figure}

The correspondence between the -0.8 eV ARPES feature and the -0.96 eV STS peak, which has been associated with the O$_{\delta}$, is reinforced by its doping dependence. A typical doping evolution is given in Fig. \ref{x}a, where the EDCs of UD70, OP90 and OD71 samples are compared. The intensity of the -0.8 eV peak increases with the doping x. In order to obtain quantitative information, we extracted the weight of the -0.8 eV peak using the following procedure, illustrated in the inset of Fig. \ref{x}a: after normalization of each EDC to the total spectral weight of the -8 to +0.1 eV binding energy range, we extracted a cubic polynomial background to the logarithm of the EDCs. Due to the strong tail of the valence band and the weakness of the -0.8 eV peak, the use of the logarithm of the EDCs rather than the EDCs improved the fits. Then, we converted back the background obtained into the natural scale (dotted curve) and calculate the area under the peak by subtracting the two curves in the -1.05 to -0.55 eV range. The errors have been estimated by evaluating how the background varies with different fit parameters and by considering the signal/noise ratio. Figure \ref{x}b shows, as a function of doping, the results obtained by averaging the weight extracted from the EDCs measured in the 28 - 100 eV photon energy range (circles). The doping x, and thus the amount of O$_{\delta}$, has been deduced from the empirical relation \cite{Presland} between T$_c$ and x given by $T_c^{\textrm{max}}/T_c=1-82.6(x-0.16)^2$ with $T_c^{\textrm{max}}$ = 95 K (plain curve). The weight extracted is consistent with the relation proportional to x (dashed line) expected for the interpretation of the -0.8 eV local state in terms of O$_{\delta}$. 

%\newpage

\begin{figure}[htbp]
\begin{center}
\includegraphics[width=8cm]{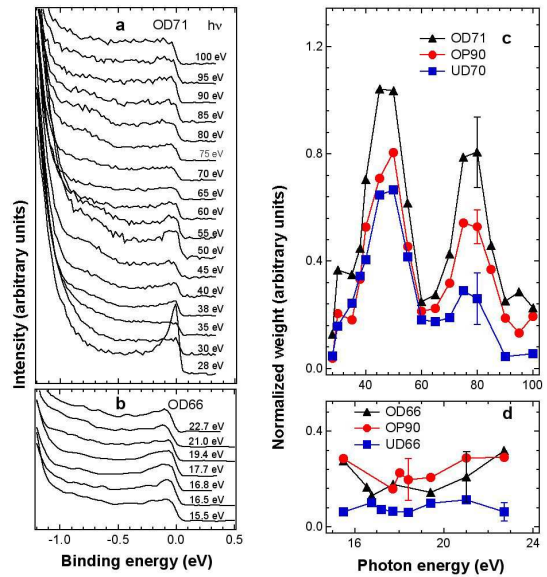}
\caption{\label{hv}(Color online) The k$_F$-centered EDCs obtained at the nodal direction are given in a) for an OD71 sample using the high photon energy range (28 - 100 eV) and in b) for an OD66 sample using the low photon energy range (15 - 23 eV). The corresponding weight of the -0.8 eV peak is given in panels c) and d) for the high and low photon energy ranges, respectively.}
\end{center}
\end{figure}

We now ask the question: How does the O$_{\delta}$ couple to the in-plane electronic properties? We investigated the electronic character of the -0.8 eV peak by performing ARPES measurements in a wide photon energy range, which are summarized in Fig. \ref{hv}. The k$_F$-centered EDCs  along the nodal direction are shown in panels a and b. While the impurity state peak does not show any dispersion, its amplitude exhibits some photon energy dependence. Interestingly, the peak is particularly intense and well defined for photon energies around 45-50 and 75-80 eV. Even though the situation is less clear for the low photon energy range (15-23 eV), a change of slope in the EDCs is clearly seen in the -1 to -0.5 eV range, as illustrated in Figs. \ref{hv}b. Using the procedure described above, we extracted the weight associated with the -0.8 eV peak for a wide range of photon energy. Figures \ref{hv}c and d show, for the 15-23 eV and 28-100 eV photon energy ranges, respectively, the results obtained for UD, OP and OD samples. The average error bars are also indicated on these figures. The weight extracted is consistent with the precedent observations. Hence, sharp resonances are observed around 50 eV and 75 eV. As for the low energy range, no resonance can be defined, within the error bars. 

%\newpage

\begin{figure}[htbp]
\begin{center}
\includegraphics[width=8cm]{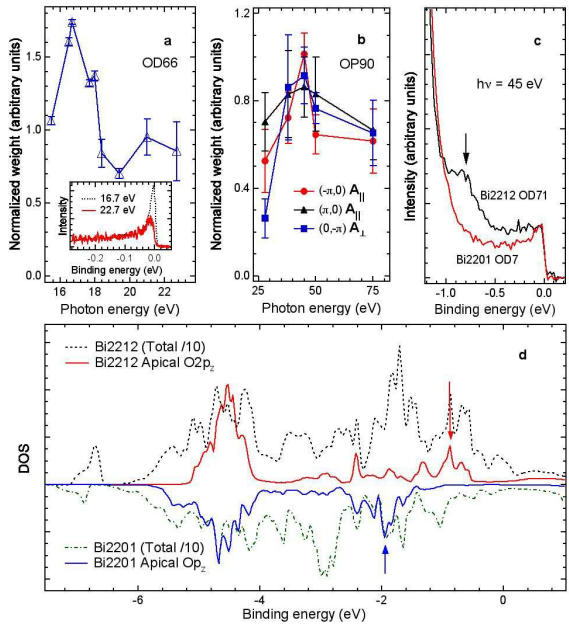}
\caption{\label{calc}(Color online) a) Photon energy dependence of the nodal quasi-particule peak weight. Inset: Comparison between the nodal k$_F$ EDCs obtained at 16.7 (dotted line) and 22.7 eV (plain line). b) Polarization dependence of the -0.8 eV spectral weight. c) Comparison between the Bi2201 (OD7) and Bi2212 (OD71) M-centered EDCs. d) Calculated total and apical oxygen O2p$_z$ density of states for the Bi2201 and Bi2212 compounds.}
\end{center}
\end{figure}

The -0.8 eV peak resonance around 75 eV suggests that this peak is, surprisingly, related to copper ions. It is well known that the Cu3p core levels locate around -75 eV \cite{Iwan,Thuler,Wendin,Shen}, and we observed them at -76.7 and -75.1 eV in our OD71 sample. Thus, one can expect that such photon energy would enhance the direct transition Cu3p$^6$3d$^9$+h$\nu$$\rightarrow$Cu3p$^6$3d$^8$+e$^-$ due to an interference with a super Coster-Kronig Auger decay process described by Cu3p$^6$3d$^9$+h$\nu$$\rightarrow$Cu3p$^5$3d$^{10}$$\rightarrow$Cu3p$^6$3d$^8$+e$^-$ \cite{Davis}. It is well established that Cu3p$^6$3d$^8$ represents a satellite state ($\sim$ -12.5 eV) of the valence band, as observed in \lsco\xspace and \ybco\xspace\cite{Shen}. In our OD71 sample, this satellite is observed at -12.3 eV and we checked that it is also strongly enhanced at a photon energy of 75 eV as compared to 70 eV. Contrary to the 75 eV resonance, the resonance around 50 eV is more difficult to interpret in terms of a copper resonance. However, previous ARPES studies of Bi2212 revealed a resonance of the anti-bounding component of the bilayer hydridization around 50 eV \cite{Kordyuk,Borisenko}. Thus, it natural to correlate the 50 eV resonance observed for the -0.8 eV peak and the near E$_F$ anti-bounding band. Remarkably, we observed no indication of any resonance associated directly with oxygen, as one would expect. For example, such a resonance, reinforced by quantum interference with the O2s$\rightarrow$O2p transition (around 17-18 eV \cite{Meyer}), enhances the near-$E_F$ band along the nodal direction, as shown in Fig. \ref{calc}a. However, it has no sizable effect on the -0.8 eV peak intensity (see Figs. \ref{hv}e and g). This indicates that the local state observed involves only oxygens with filled 2s and 2p shells (O$^{-2}$). 

%\newpage

Even though the localized state observed at -0.8 eV is induced by the O$_{\delta}$, its precise origin remains unclear. The most natural explanation of our results is to interpret the -0.8 eV peak as a O$_\delta$ level. A recent DFT study has established that the O$_\delta$ site is most likely located between the SrO and BiO layers \cite{He}, with a small ab displacement as compared to the STS results \cite{Mcelroy}. It has been concluded from LDA calculations performed on the relaxed structure obtained by DFT that the feature observed by STS, and thus by ARPES, comes from unhybridized dopant O2p$_z$ states \cite{He}. However, this hypothesis cannot explain the photon energy resonances observed by ARPES. In an alternative scenario, the -0.8 eV peak can be interpreted as O$_\delta$ states strongly mixed with their environment. The O$_\delta$ is likely to be mixed with Cu through the apical oxygen O2p$_z$ and Cu3d$_{3z^2-r^2}$ A$_{1g}$ orbitals. The absence of strong polarization dependence illustrated in Fig. \ref{calc}b is consistent with this scenario and disfavors the possibility of a mixing of the -0.8 eV local state with the Zhang-Rice singlet (Cu3d$_{x^2-y^2}$ and O2p$_{x,y}$ bands) \cite{Zhang}. In order to investigate further this second scenario, the total density of states (DOS) and the DOS of the apical oxygen O2p$_z$ band have been calculated for both Bi2201 and Bi2212 using LDA+U. As indicated by the results, given in Fig. \ref{calc}d, the DOS of the apical oxygen O2p$_z$ band in Bi2212 exhibits a significant feature around -1 eV (indicated by arrows), which is moved by about 1 eV towards the higher binding energies in Bi2201. As a consequence, the absence of the -0.8 eV peak in the experimental overdoped Bi2201 EDC (OD7) illustrated in Fig. \ref{calc}c suggests that this peak comes from spectral weight pulled out from the valence band, following the mixing of the O$_{\delta}$ with the apical oxygen O2p$_z$ band, which is itself hybridized to the Cu3d$_{3z^2-r^2}$ band. Moreover, the complementarity between ARPES and STS allows us to predict that the -0.96 eV peak observed by STS will not be observed in Bi2201.

Among the possible consequences of the second scenario proposed in this letter, a local modification of the Cu-Cu second nearest neighbor hopping term t$^{\prime}$ is expected since the Cu3d$_{3z^2-r^2}$ and apical O2p$_z$ orbitals related to this parameter are themselves strongly perturbed in the vicinity of the impurity site \cite{Pavarini,Eisaki,Yin}. This leads to correlations between the O$_\delta$ distribution and the CuO$_2$ electronic properties, as evidenced by the gap magnitude inhomogeneities reported by STS \cite{Mcelroy}. However, it is not clear that the observed weight of the -0.8 eV resonance can account for all oxygen dopants. Actually, a careful counting of the impurity sites observed by STS indicates that only a fraction of the dopants ($\sim$0.5) are detected \cite{Mcelroy}, suggesting the possibility of additional O$_{\delta}$ non-equivalent sites for which the ARPES and STS techniques are not sensitive. For this reason, further investigations are needed for a full understanding of the local impact of the dopants on the CuO$_2$ plane electronic properties and on high-temperature superconductivity. 

\begin{acknowledgments}
We thank J.C. Davis, P. J. Hirschfeld, P.A. Lee, and F.C. Zhang for valuable 
discussions and suggestions.
This work is supported by NSF DMR-0353108, DOE DE-FG02-99ER45747. This work is based upon research conducted at the National Synchrotron Light Source supported by
DOE DE-AC02-98CH10886, and at the Advanced Light Source supported by DOE DE-AC03-76SF00098. 
\end{acknowledgments}

\bibliography{biblio_ens}

\end{document}